# A Framework for the Implementation of Industrial Automation Systems Based on PLCs

Kleanthis Thramboulidis
Electrical and Computer Engineering
University of Patras, Greece
*thrambo@ece.upatras.gr*

*Abstract*—**Industrial automation systems (IASs) are traditionally developed using a sequential approach where the automation software, which is commonly based on the IEC 61131 languages, is developed when the design and in many cases the implementation of mechanical parts have been completed. However, it is claimed that this approach does not lead to the optimal system design and that the IEC 61131 does not meet new challenges in this domain. The increased complexity of these systems and the requirement for improved product and process quality attributes impose the need for exploiting state of the art trends in Mechatronics and Software Engineering. In this paper, a system engineering process based on the new version of IEC 61131, which supports Object Orientation, is presented. This approach adopts the synergistic integration of the constituent parts of mechatronic systems at the Mechatronic component level. Synergistic integration, component based development as well as key concepts of service and object orientation are properly integrated to define a solid framework for addressing the needs of today's complex industrial automation systems. SysML and UML are utilized to introduce a higher layer of abstraction in the design space of IAS and Internet of Things (IoT) is considered as an enabling technology for the integration of cyber and cyber-physical components of the system, bringing into the industrial automation domain the benefits of these technologies.**

## I. INTRODUCTION

### A. The System Development Process

INDUSTRIAL automation systems perform physical and computational processes. The physical processes, which are executed by the physical plant, are monitored and controlled by computational processes which are performed by networks of embedded computation nodes. These systems are developed following the traditional approach, according to which their constituent parts, i.e., mechanics, electronics, and software, are developed sequentially and independently, and then are integrated to compose the final system. The traditional approach, which is also adopted by the majority of published work regarding the software part of industrial automation systems, does not lead to optimal system behavior, since it does not adopt a system level approach and does not properly focus on the interactions among the constituent components [1]. Design decisions are taken from the point of view of a single domain so they do not take into account their impact in other domains [2]. Moreover, today's requirements for flexibility, reduced energy consumption, environmental sustainability, improved product quality and reduced product cost, generate new challenges in the development of IASs. These challenges cannot be addressed by the traditional approach [3][4]. A mechatronic system, such as IAS, requires a multidisciplinary approach for its modeling, design, development and implementation [5][6], and the concept of the Mechatronic component plays a key role in this direction [7].

Furthermore, developers following the traditional approach use different tools, different notations and different methodologies for the modeling and implementation of the various discipline parts of the system. Every discipline has its own approaches; an integrated framework for the construction of mechatronic systems is missing [8]. This makes the analysis of the whole system during the development process a very difficult and in many cases an impossible task [7]. As a result, there has been an increasing interest of the research community in mechatronics over the last years. New methodologies have been proposed to address the challenges in the mechatronic system development. Current trends in this domain propose the synergistic integration of the various discipline parts of the mechatronic system [1]. Research results, e.g., [9], can be properly exploited in the development process of IASs.

### B. The Cyber part

The software part of industrial automation systems is largely based on Programmable Logic Controllers (PLCs), which are usually programmed using the languages defined by the IEC 61131 standard [10]. Control engineers have widely adopted the IEC 61131 standard [11] to specify the software part of their systems, mainly when programmable logic controllers (PLCs) are used [10][12]. IEC 61131, which was first published in 1992, defines a model and a set of programming languages for the development of industrial automation software and is considered as one of the most important standards in industrial automation [13]. However, the standard has been criticized the past few years for not addressing any more the requirements of today's complex industrial automation systems and not being compliant with state of the art software engineering practices [14][15]. As claimed in [10], the new challenges of widely distributed automation systems are not addressed by the standard. Most of the main requirements for future automation systems, are not supported by the industrial solutions available on the market [16].

Analogous requirements have been successfully addressed in the general purpose computing domain by exploiting state-of-the-art technologies of Software Engineering such as Object-Orientation (OO), component based development (CBD) and service oriented computing (SOC), but first of all by raising the level of abstraction in the development process through the Model Driven Engineering (MDE) paradigm. The exploitation of these technologies in the industrial automation domain is a challenge for academy and industry. An early example of this trend is the IEC 61499 standard, which was proposed as an extension of the 61131 FB model to exploit the benefits of OO. However, even its life is almost 10 years it has not been adopted by industry [10][17]. The fail of 61499 to meet its objectives is one of the reasons for triggering the discussions on an extension of 61131 to provide support for OO. In 2013 the third edition of IEC 61131-3 – Programming Languages, version FDIS (Final Draft International Standard) of 2012 was approved as International Standard and version 3.0 of IEC 61131 is officially available as International Standard by IEC. As argued in [18], this version of 61131 provides a better support to OO compared to 61499 through the use of constructs of interface, class, OO FB and the implementation of concepts such as inheritance and implements. However, both 61499 and version 3.0 of 61131 have two common drawbacks. Firstly, they do not base their OO support to the existing support for OO that is provided by version 2.0 of 61131 [19][10], and secondly, they do not extend the partial support for MDE provided by version 2.0 of 61131. These relations are shown in Figure 1. Moreover, researchers are already working on the direction of extending IEC 61131 to provide support for components based development and service orientation [20][21]. It is expected that the new version of 61131 will dominate in the next generation of IASs if properly extended to exploit these trends in software engineering.

### C. Contribution and organization of the paper

To meet the above requirements we have adopted the 3+1 SysML-view model that is an implementation of the Model Integrated Mechatronics (MIM) paradigm [4] based on the System Modeling Language (SysML) and we adapt it to the industrial automation domain. MIM focuses on the synergistic integration of the discipline parts and defines the system as a composition of Mechatronic, i.e., cyber-physical, components. Internet of Things (IoT) [22] is proposed to be used as "glue" to integrate the constituent components of the Mechatronic system as far as it regards their cyber-interfaces. SOA-based architectures for the IoT middleware are evolving, e.g., [23], and will play a key role in this domain. For the software part of the mechatronic component we use the languages of version 3.0 of IEC 61131 that provide support for OO. However, FBD and ST of version 2.0 can also be used since they provide a partial support for OO. Even though IEC 61131 has a life of 20 years it is still widely used for the implementation of industrial automation software that has real-time constraints, even in the case that new technologies such as agents or service orientation are utilized in this domain [24].

We define the development process of the software part of the industrial automation system to properly fit in a system's development process. We adopt MDE and use SysML/UML to define a higher layer of abstraction in the design of industrial automation software compared to the one provided by IEC 61131. The automatic transformation from this higher abstraction layer to 61131 is not a subject of this paper. In this paper, we present and describe our approach for modeling IASs using SysML/UML and compare it to the one presented in [10] which is based on Petri nets. To facilitate this comparison we use as case study the one used in [10]. The merging of the two approaches is also considered and discussed towards a more rigorous development process. UML and SysML are used by several research groups in 61131 based systems. Authors in [25] present a methodology to evaluate the benefits and usability of different modeling notations, including UML, SysML and IEC 61131, for open loop control in automation systems.

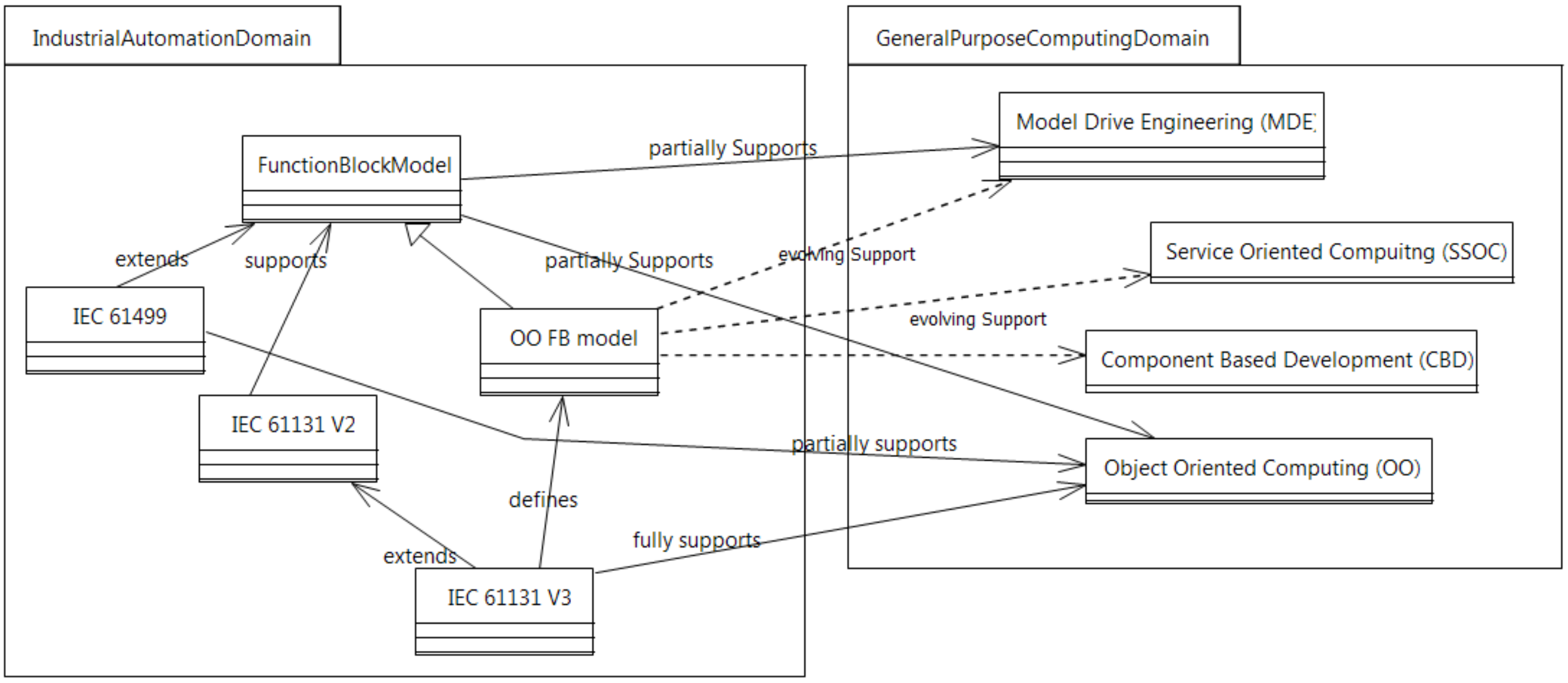

Fig.1 Relations among industrial automation standards on the Function Block model and SE technologies.

The remainder of this paper is organized as follows. In Section 2, work related with the above presented framework of technologies is discussed. In Section 3, the system that is used as case study in this paper is described and the proposed system development process is presented. Section 4 presents the proposed approach for the development of the IEC 61131-based software. In Section 5, the proposed approach is discussed along with the one presented in [10] to highlight the pros and cons of both approaches towards their integration. Finally, the paper is concluded in the last section.

## II. Related Work

Trends in Mechatronic system development as well as in Software Engineering, such as Object Orientation and model-driven development, have motivated researchers to find alternatives to exploit their benefits in 61131-based systems. Authors in [18] have examined alternatives for extending the 61131 model to support OO and presented the benefits of such an extension. To this direction is also the work of the TC65 working group of IEC in the context of version 3.0 of 61131 that has as its main feature new constructs to support the OO paradigm in this domain.

Basile *et al.* claim in [10] that the languages of the IEC 61131 standard are not ready to meet the new challenges of widely distributed automation systems; these languages are considered not efficient to describe sequential behavior and concurrency when the system complexity grows. To address these requirements authors adopt the service-oriented paradigm, an event-based execution model, formal methods and supervisory control theory. They describe an approach for extending the IEC 61131 model to support, according to authors, an event-based execution order in a similar way with the IEC 61499 standard. Authors propose the use of two types of FBs. Device FBs (DFBs) are used to implement the basic functionality that requires access to I/O filed signals. Operation FBs (OFBs) are used to implement operations that use the functionalities provided by DFBs to perform certain working cycles such as filling of a silo and transfer of a pallet. Authors claim that the event-based execution order allows the FB model to fulfil the service requests associated with input events. Supervisory control is used to solve the problem of programming the concurrent FB behaviors so as to satisfy desired sequencing and logical constraints. This model facilitates, according to the authors, the development of distributed automation systems avoiding the disadvantages of the IEC 61499 standard. However, their approach and mainly their example implementation through which they describe their approach is not distributed not even contain concurrently executed computational behaviors; the example application is developed based on the scan cycle model that is executed on one PLC as a sequence of FB calls inside a cycle. This work is extensively discussed in this paper since parts of this work can be integrated with the work presented in this paper to form a rigorous development process for IAS.

Model-Driven Engineering has motivated researchers in the industrial automation domain to look for alternatives to exploit its benefits in the IEC 61131 based systems. Most of these works, as discussed in [19], propose a higher layer of abstraction in the design of the 61131 software by exploiting UML, e.g., [26][27][28][29] or SysML, e.g., [30]. However, as claimed in [19], these works do not exploit the OO aspects of IEC 61131 and this has resulted into inefficient mappings of UML and SysML constructs to IEC 61131 constructs. Moreover, as claimed in [18] none of the current works address the problem of absence of constructs that will allow a more efficient application of the MDD paradigm based on the IEC 61131 languages.

Witsch *et al.* are adopting in [31] the OO extension of IEC 61131 and describe a mapping of 61131 constructs to UML constructs. Mapping rules between these constructs are defined. The class diagram is used to capture the structure and activity and state charts are used to capture the behaviour of the software that is to be implemented based on the OO 61131 model. CoDeSys v3.0 is used as a prototyping tool for the proposed approach. In [29], authors present an approach that is based on a UML profile for the model driven development of 61131 and a tool that has been developed to support this approach. Authors in [32] present an extension of SysML block definition diagram to support architectural design of software product lines. Jamro *et al.* describe in [33] an MDD approach that utilizes four SysML diagrams, i.e., requirements diagram, package diagram, block definition diagram, and state machine diagram. The requirements diagram is used according to authors to "present requirements for POUs of IEC 61131-3 control software." The package diagram is used "to represent controllers and task assignments." Two models, i.e., the resources and the task one, are constructed using the package diagram. The task model is used to capture the assignment of POUs to tasks. The block definition diagram and the state machine diagram are used to model POUs adopting the approach proposed in [30]. POUs are represented using the construct of block and inputs and outputs of POUs are represented using the flow port. It should be noted that the package in SysML is the basic unit of partitioning of the system model into logical groupings to minimize circular dependencies among them. SysML also introduces the allocation relationship that can be used to allocate a set of model elements to another, such as allocating behavior to structure or allocating logical to physical components. The deployment of software to hardware can also be represented in the SysML internal block diagram. Flow ports as well as flow specifications that were used to specify the flow ports have been deprecated in version 1.3 of SysML.

From the above works it is evident that researchers have focused on the exploitation of MDD in parallel with the adoption of the OO paradigm. However, the new version of 61131 does not make any contribution towards a better support for MDD [18].

In the past few years, new trends in Mechatronics have motivated researchers in the industrial automation domain to look for alternatives to exploit these trends in industrial automation, e.g., [34][35], and more specifically when developing systems based on IEC 61131, e.g., [36] and [37]. Authors in [34] consider the system composed of three levels: the system, the modules and the components level. However, they do not clearly define the semantics of these three levels not even the level of the synergistic integration of the three disciplines. Moreover, they criticize the V model for introducing the control system after the development of components to cover the gap of the mechanics. They base their decision to adopt the W model on this critic. They claim that the W model provides a virtual system integration, i.e., a high-level system model simulation. However, as it is evident from the MTS-V model [4] and the presented in this paper approach this critique is not correct. In MDE the constructed models are executable and the verification of the design is performed before the development of the system's components.

To the best of our knowledge there is no other work that describes a development process for industrial automation systems that, a) utilizes a higher layer of abstraction in the design space compared to the one provided by the FBD language of the new version of IEC61131, and b) exploits the benefits of model driven engineering at system level and the new trends in Mechatronics such as synergistic integration at the component level.

## III. Modeling The System

Industrial automation systems are composed of the physical plant that performs the physical processes and networks of embedded computers that perform the computational processes required to monitor and control the physical processes. Computational processes accept inputs from the physical processes and calculate the outputs required to affect these. In this section, a) we describe the example system that is used as a case study in this paper, and b) we briefly describe the system level development approach that is proposed in this paper. We focus on the cyber part, which is the main subject of this paper, in the next section.

### A. *The liqueur plant example system*

The example application used in [10] is adopted as case study to demonstrate the features and capabilities of the approach presented in this work. We assume that the system under development, i.e., the target system, is a plant for generating two types of liqueurs, type A and type B. Figure 2 presents the mechanical part of the plant, i.e., the physical part of the target system that performs the physical processes. As shown in Figure 2 the example plant is composed of four silos connected by a pipe. Each silo *i* has an input valve *INi* and an output valve *OUTi* through which is cyclically filled and emptied with liquid. It also has a sensor *Ei* for the lower level and a sensor *Fi* for the upper level. Two silos (2 and 4) have a resistance *Ri* to heat the liquid and a sensor *Ti* to monitor the temperature. Two silos (3 and 4) have a mixer *Mi* to mix the content of the silo. Simplified descriptions of the two processes that are executed in the plant are assumed. Silos S1 and S4 are used for the production of liqueur A. Raw liquid undergoes a basic process in S1 and then it is poured into S4 where it is further processed, i.e., it is heated and then mixed. Silos S2 and S3 are used for the production of liqueur B. Raw liquid is heated in S2 until a given temperature is reached and then it is transferred to S3 where it is mixed for a given time. The two processes are independent and can be executed in parallel. However, since our example plant uses the same pipe for liquid transfer between silos, the two processes should be synchronized. There is one more constraint regarding power consumption. Mixing the liquid in silos S3 and S4 at the same time is not permitted.

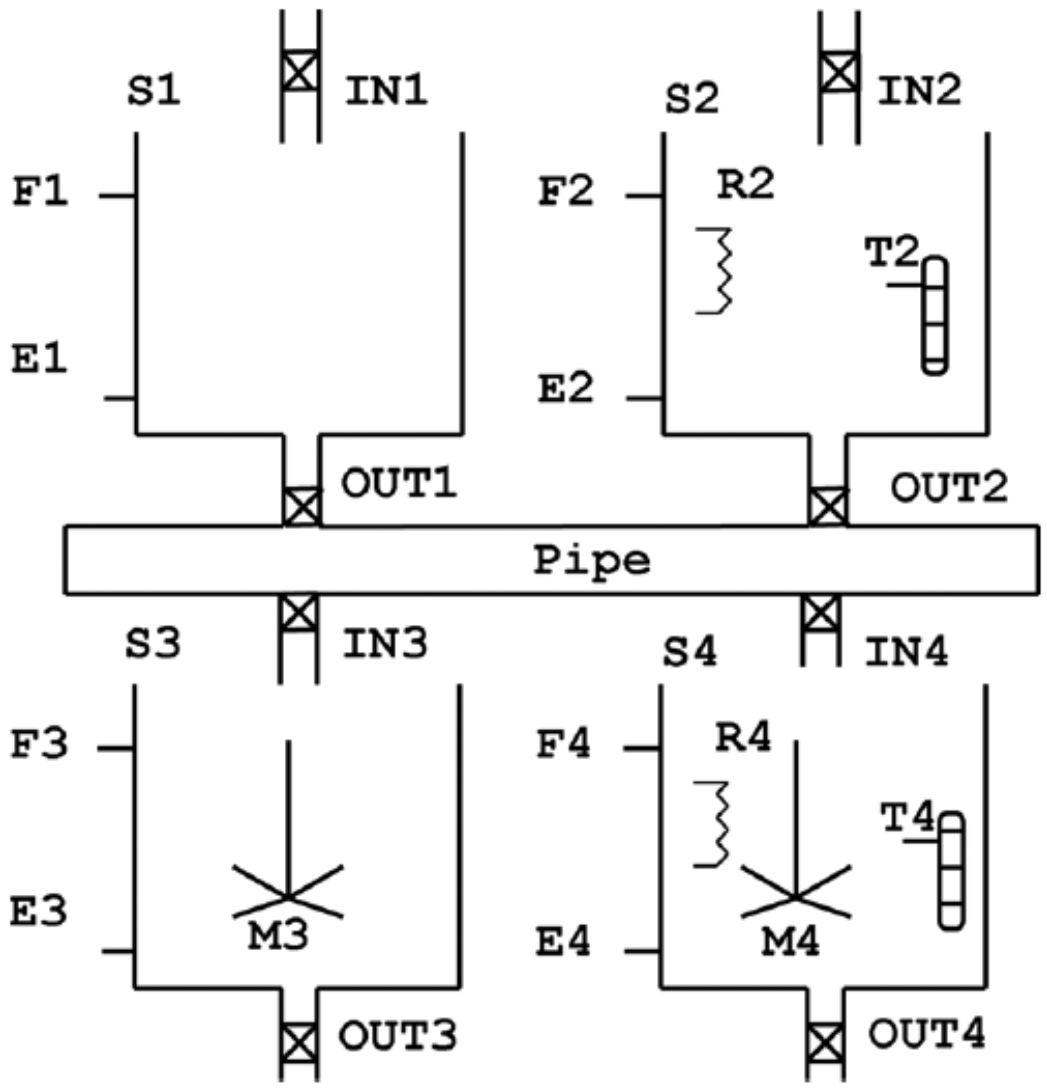

Fig. 2. The physical plant example system [10].

### B. *The System View*

We adopt MIM [1] and we model the system at the Mechatronic layer using SysML [4]. This model is next projected at the three discipline layers, i.e., mechanics, electronics and software, to get the corresponding views (M-view, E-view and S-view) as shown in figure 3. The system developer is working on the SysML-view while engineers of the three disciplines work on the corresponding views [4]. Model elements represent, among others, cyber, cyber-physical, and physical components that constitute the target system as well as relations among them. View elements are the representation of model elements in diagrams of the various views.

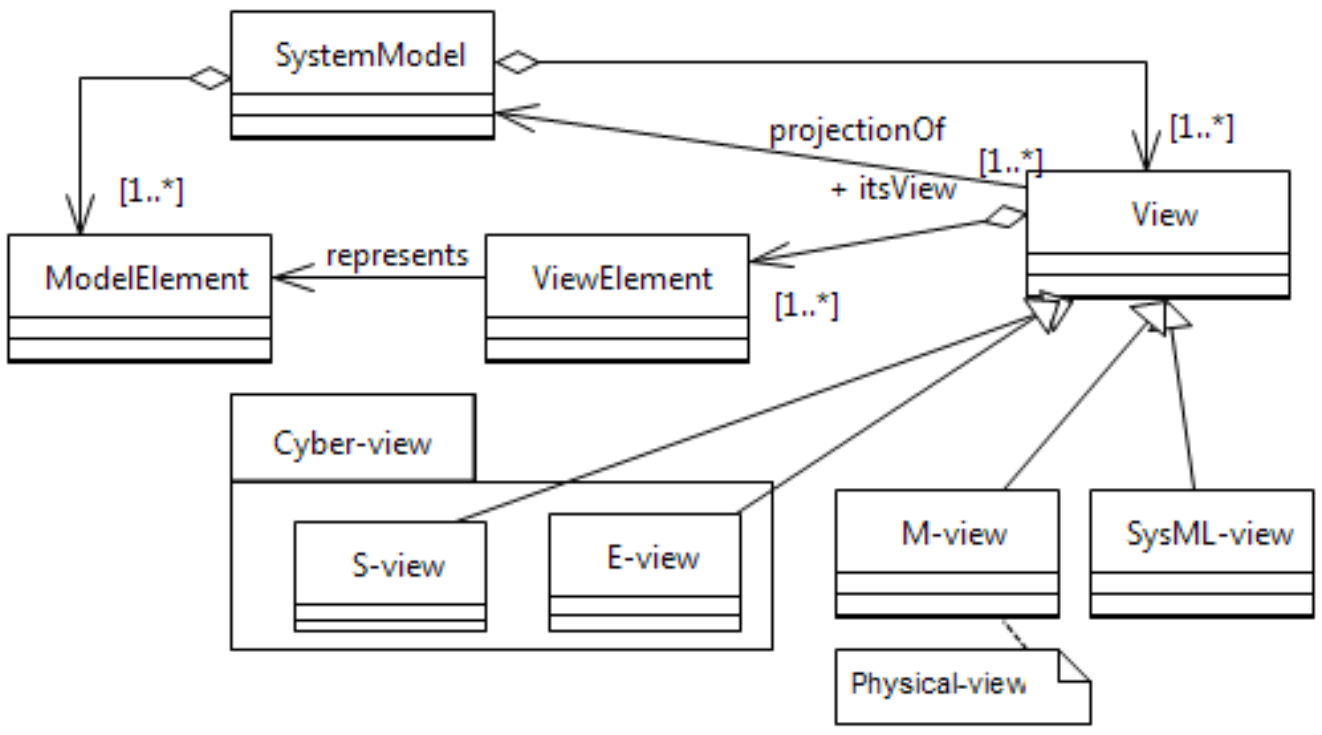

Fig. 3. Relations between the system model and the various views of the system.

Following the MIM paradigm we start the development process with requirements modeling at system level. The development process is based on the MTS V-model that is a V-model adopted to the requirements of mechatronic systems [4]; SysML is used as modeling language at this level. The SysML requirements diagram and essential use cases are used to capture the required by the system functionality in terms of system responsibilities, as well as required QOS characteristics. Based on requirements the architecture of the system is defined at the system layer of the MIM architecture as a composition of Mechatronic Components (MTCs). The developer has to split the system level functionality and assign chunks of functionality to system level components. Moreover, the collaboration of components should be defined so as to obtain the system level functionality that has been captured in SysML requirements diagrams. Chunks of system level functionality that include physical processing are allocated to cyber-physical components. For example, a cyber-physical component of type Silo is required for a processing on the raw liquid used for the liqueur production. This Silo cyber-physical component is composed of the mechanical silo, an embedded board that provides storing, processing and communication capabilities and a software layer that runs on it to monitor and control the mechanical silo and convert it into a smart silo. In a similar way, the liquid transfer functionality is assigned to a smart Pipe, another cyber-physical component that integrates the knowledge of performing specific operations, e.g., define the velocity of transfer, clean the mechanical pipe or acquire and release it.

The liqueur generation process is modeled as a cyber component since it captures only the coordination logic of the cyber-physical components that are used to realize the functionality described by the specific process. Figure 4 presents part of the architecture of the target system. Cyber-physical components, such as Silo, MixedHeatedSilo and Pipe, offer their functionality as services through well defined interfaces. These services may be orchestrated by cyber components, which implement plant processes, to realize the functionalities required at system level. Complex processes may be decomposed to sub-processes to handle complexity, if any, at the process level. For example, in our plant system the MixedHeatedSilo offers services such as filling, empting, heating and mixing. These services are used by the *GenLiqueurA* cyber component, along with the services offered by the Silo and Pipe components, to fulfill the requirements for liqueur A production. Web Services can be used as a means for this integration [20-21][38]. In this work we consider the Internet of Things (IoT) as the emerging technology for the integration of cyber and cyber-physical components bringing into the industrial automation domain the benefits of this technology in the form of Intranet of Things.

We have modeled MixedHeatedSilo as a primitive cyber-physical (primitiveCP) component, which means it may not be further decomposed in terms of MTCs. A primitive MTC is decomposed in terms of its three discipline parts, i.e., mechanical, electronic and software, which are concurrently developed and integrated realizing synergistic integration of the three disciplines at the component level. The bottom part of the MTS-V model captures this synergistic integration process [4]. In the case that Heater and Mixer are already existing cyber-physical components then MixedHeatedSilo should be defined as composite MTC (compositeCP) consisting of three cyber-physical components. IoT may be used even in this case as a "glue" regarding the cyber interfaces to integrate the constituent components of compositeCPs. The composition association that connects a system level component with its constituent components justifies the characterization of this component as a compositeCP component, as is the case of LiqueurPlant shown in figure 4. On the other side, the use of aggregation relationship, as in the case of process components, e.g., *GenLiqueurA*, characterizes the component as cyber component.

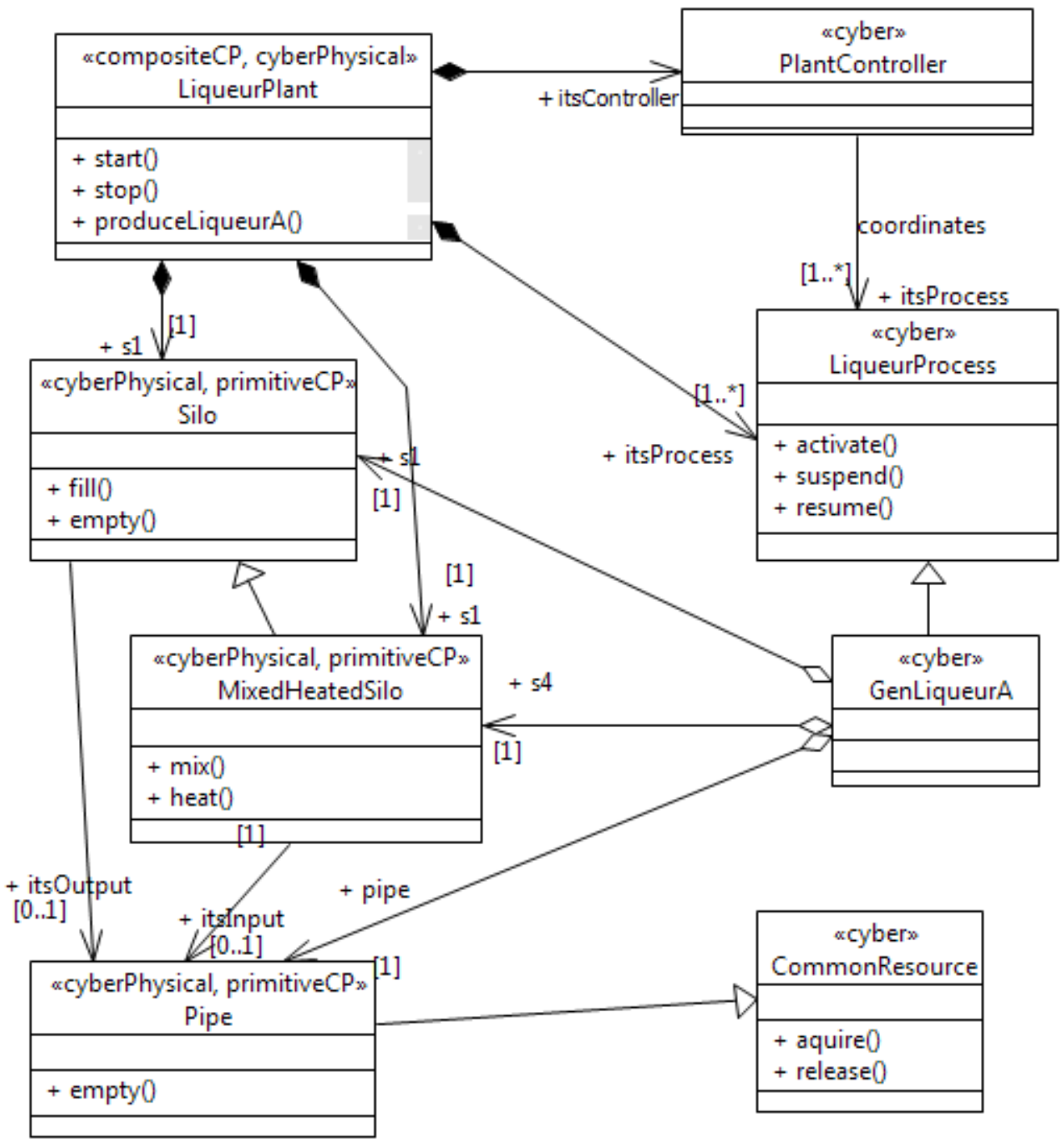

Fig. 4. The system Architecture of the liqueur plant (part).

To exploit the benefits of the service oriented paradigm we construct the «cyberPhysical» components as service providers who's services can be orchestrated by process components to realize their behavior. Services offered by the «cyberPhysical» component should be specified in the architecture diagram at the system level. The concept of port is used to represent the connection and interaction points of the component with its environment. Provided and required interfaces specify the software ports (s-ports) which have to be allocated to e-ports. Liquid input and output is performed through m-ports. Energy and signal flow is realized through e-ports. This approach compared to the one in [10], where input events of FBs are considered as service requests, exploits in a better way the service oriented paradigm. Figure 5 presents the Silo «cyberPhysical» component. There is a standard port (*processPort*) that represents the interaction point of Silo with the process cyber component, as shown in figure 5(a). An e-port is used to represent the Internet interface of the electronic part of the component, an e-port for power and two m-ports to represent the flow of liquid. The *processPort is* specified with provided (*Silo2ProcessIf*) and required (*Process2Unit ControlerIf*) interfaces as shown in figure 5(b). *Silo2ProcessIf* represents the interface that should be implemented by the process controller so as to be properly integrated with the Silo cyber-physical component. *Process2UnitControlerIf* is the interface that can be used by the process component to realize the physical process it implements.

## IV. Modeling the Cyber Part

The cyber view of the system is constructed by projecting the system model, which is expressed in SysML (see figure 4), to the cyber domain. Since at this stage of modeling we only model system level functionality and we do not assign chunks of this functionality to electronic parts, this view can also be considered as the software view (S-view) of the system. Figure 6 presents the diagram of this view that captures the structure of the software that is used to realize the process of generating liqueur A. We do not capture the electronic part of the cyber-physical system into functional diagrams since we have selected not to implement any system level functionality directly on this layer. This is why the Silo cyber component is represented only by its software constituents. However, it should be noted that even in this case the properties of the electronic part (execution platform) are of interest and should be captured in another diagram, since they greatly affect the quality of service characteristics (non-functional properties) of functionalities implemented by the software part.

In this section we focus on modeling the structure and the behavior of components of the cyber-view, which are constituent parts of cyber-physical components. For every cyber-physical component of the system level there is a cyber component in the cyber-view, e.g., the Silo «cyber» shown in figure 6 for the Silo «cyberPhysical, primitiveCP» of figure 4. We also consider the modeling of behavior of cyber components that realize plant processes, as for example the *GenLiqueurA* shown in figure 6. We briefly describe the approach presented in [10] for the modeling of these two type of components and then we describe our proposal. This approach utilizes UML/SysML to provide a more abstract and expressive design that is next automatically transformed to IEC 61131 specification applying the model driven engineering paradigm. The adoption of the new version of IEC 61131 allows for a more straightforward mapping of the SysML/UML design specs to the implementation language constructs. However, a mapping to the widely used today IEC 61131 is also possible exploiting the already existing OO support that is provided by version 2.0 of the standard.

An integration of the two approaches, as for example the use of supervisory theory and petri-nets used in [10], will bring the benefits of these technologies to our framework and will result into a more robust infrastructure for the development of IAs.

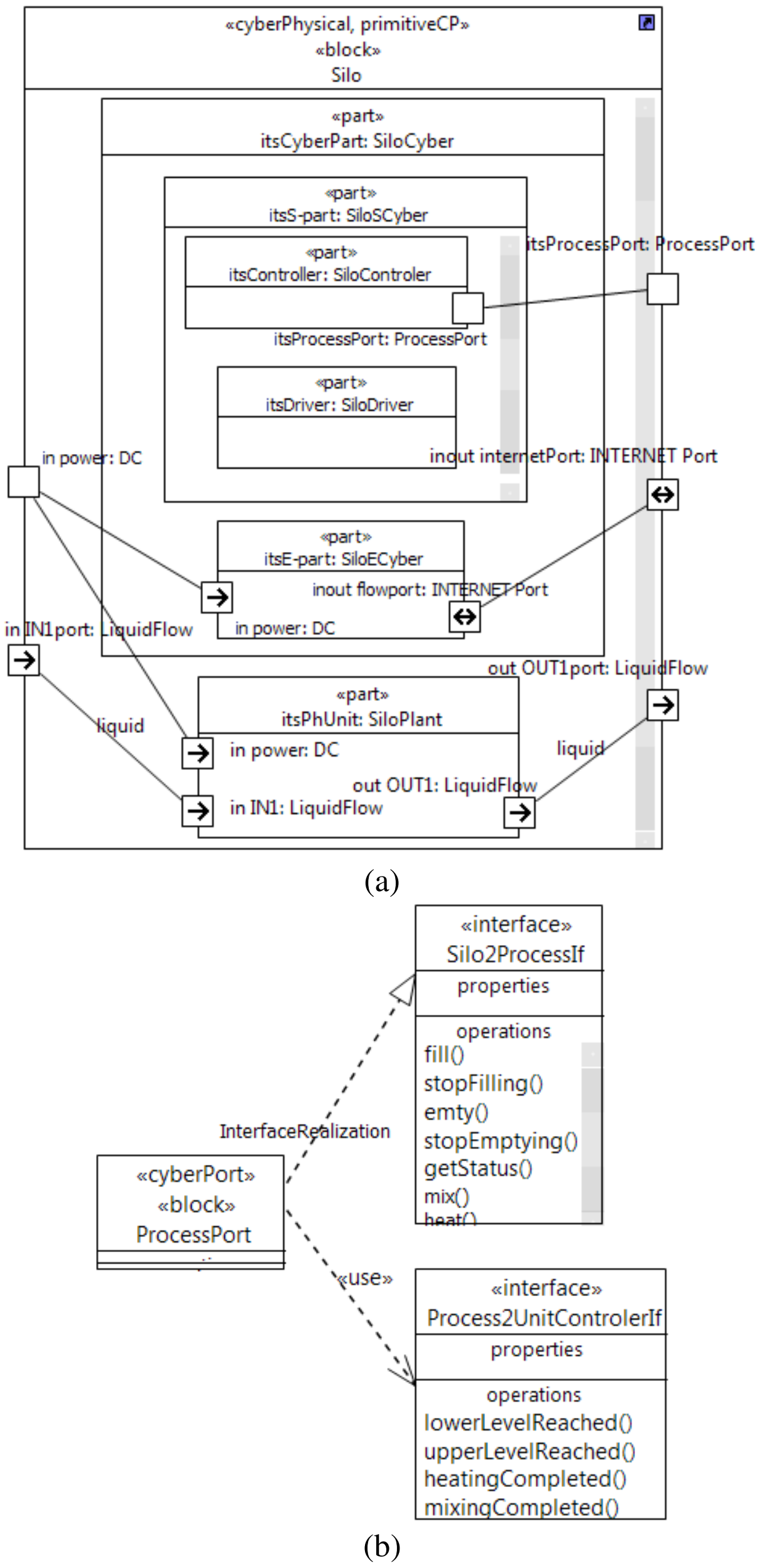

Fig. 5. (a) The Silo «cyberPhysical» component. (b) Provided and required interfaces of the cyber port *ProcessPort* of Silo.

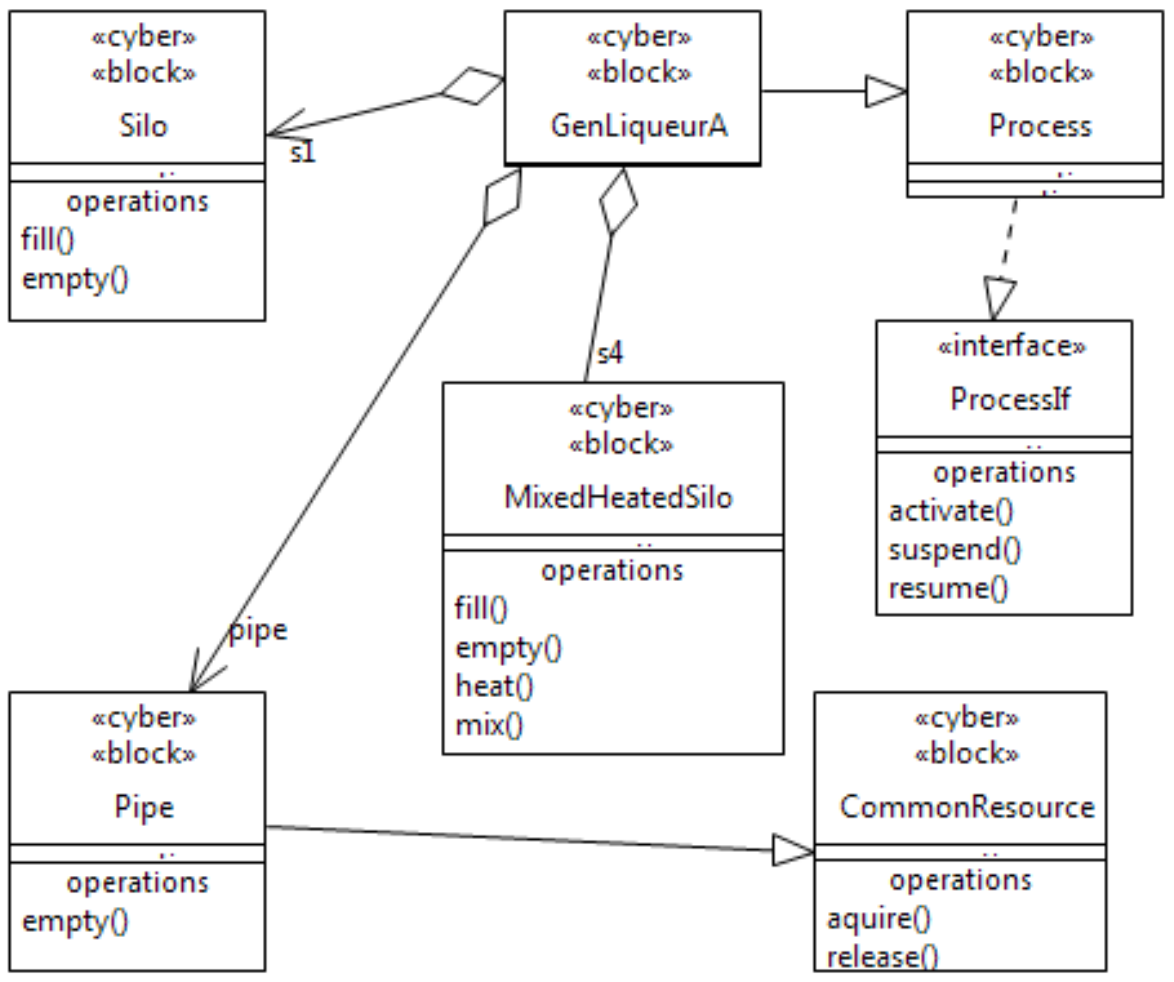

Fig. 6. The architecture of the cyber part of the system (part).

## A. *Device FB, Operation FB and Petri Nets*

For modeling the structure and behavior of the software that is closely related with Physical objects of the Plant, such as the silo in the example application, authors in [10] propose the use of specific classes, i.e., the Device FB (DFB) and the Operation FB (OFB), to increase, as they claim, reusability and adaptability. DFBs are used to implement basic functionalities that require access to I/O field signals. These functionalities are used by OFBs to implement specific working cycles, such as filling and emptying a silo. The FILLING OFB is defined, for example, to implement the whole sequence of silo filling and the conversion of sensor measures. In figure 7 we present the proposed in [10] structure for the cyber part of Silo.

Authors in [10] do not clearly state if and where they capture the state of the physical object. A pointer is used to link the OFB with the corresponding DFB. This allows, according to the authors, the use of methods or properties of the DFB inside the OFB by the standard dot notation and in this way to use all the benefits of OOP. After the definition of the FBs authors proceed to the definition of the body of the automation program. They propose an approach, which they call event-based, according to which the body consists of FB callings by means of methods. Authors argue that SFC, which is widely used for this purpose, is not sufficient when the system complexity grows. To address this problem they use Petri nets (PNs) [39][40]. They propose the modeling of sequences of a desired behavior, such as the process of producing liqueur of type A (*GenLiqeuerA*) by means of a PN, which they call PN controller. The role of the PN controller is to call FBs. It sends to FBs the order to start a certain service and it receives from FBs the event of service completion. They implemented the orders send to FBs as FB, i.e., method, calls. The two PN controllers, which are defined to model the corresponding processes of the example plant, are next implemented in the body of the automation program, which is executed based on the scan cycle model. For handling the constraints on using the common resources, such as the pipe and power, they exploit supervisory theory [41]. They define a supervisor which acts on the events generated by the PN controllers and by FBs, to force the PN controllers to satisfy the constraints. They model the supervisor using PNs. The PN model of the supervisor is integrated with the PN models of the two controllers and the resulting PN is realized by the program's body which is cyclically executed on the PLC. The IEC 61131 program (PROGRAM SILOS CORDINATION), they have developed, instantiates the DFBs, the OFBs and connects them properly.

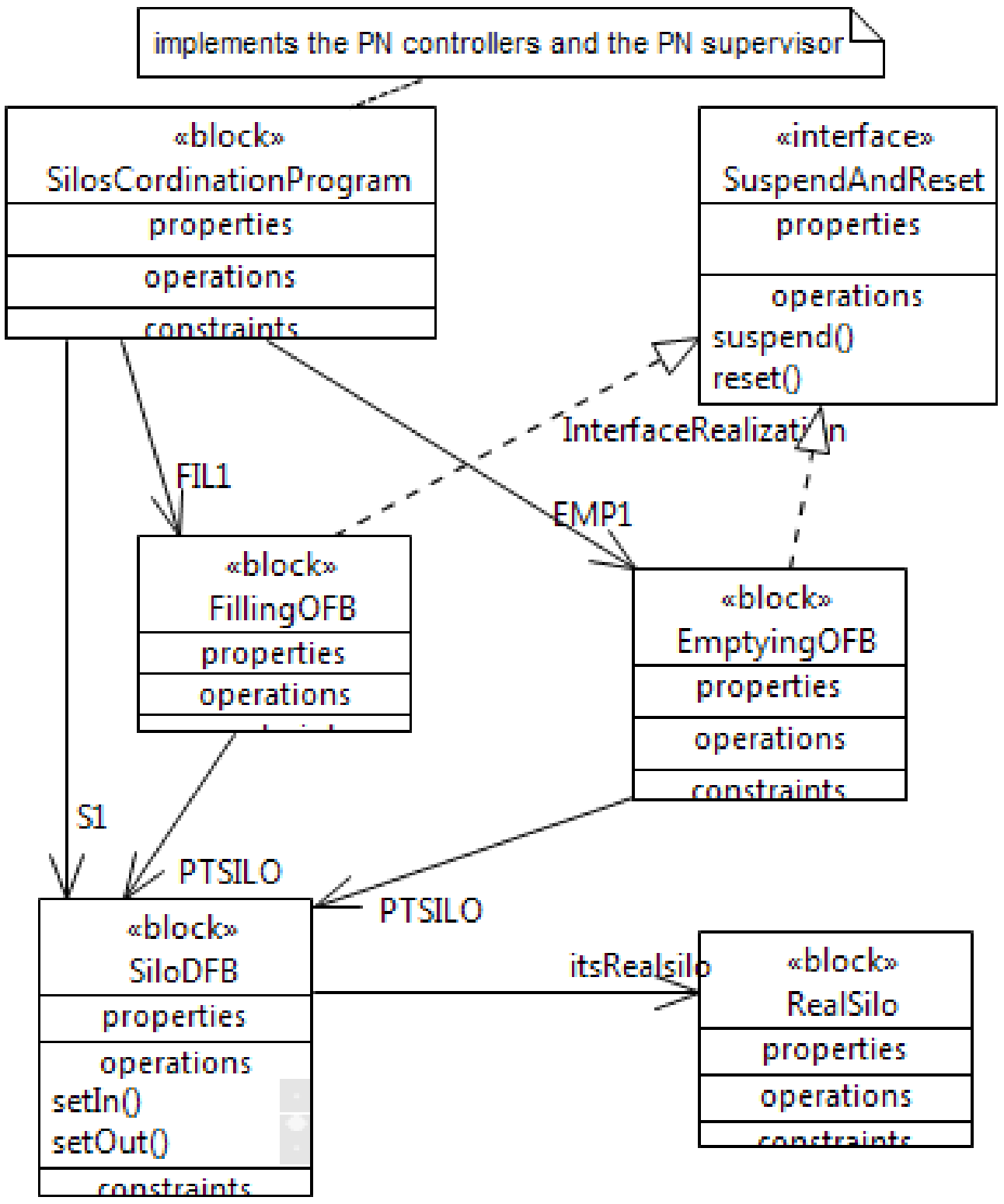

Fig. 7. Part of the basic structure of the example application as proposed and implemented in [10].

## B. *The proposed design approach*

The initial architecture of the software view (S-view) of the automation systems results from the SysML system model. This architecture should be further refined to be mapped to the implementation platform that is the OO 61131. For this refinement we use specific UML design constructs that have been defined by a UML profile taking into account the implementation platform capabilities and constraints. In this way UML is used as a means to define a domain specific modeling language since it provides constructs directly related with the key concepts of the industrial automation domain. This design will next be mapped to 61131 constructs. In this paper a manual transformation has been performed. The automation of this transformation process is under development.

As shown in figure 5a the SmartSilo «cyberPhysical» component has been assigned behavior that can be partially executed by the Silo «physical» and the Silo «cyber». The Silo

«cyber» should be assigned the behavior of controlling the Silo «physical» to perform the operations fill() and empty(), which have been assigned to Silo «cyberPhysical». It should also implement the functionality required to interface the controller logic with the physical object. To increase modularity and reusability we separate these two behaviors into two FBs and consequently into two classes. FBs that capture the controller logic of the physical object are modeled using the «cpController» stereotype and implement the provided at the cyber level interface of the cyber-physical component, which is the *SmartSiloIf* shown in figure 8(a). The FB of type «cpController» communicates with the real object, e.g., physical Silo, through another FB who's type encapsulates the details of interaction of the physical unit with the software world. This FB is modeled as a «Driver» stereotype, as shown in figure 8(a), and is actually the proxy of the physical object into the software domain. The interaction between the controller and the driver is specified in terms of provided and required interfaces as shown in figure 8(b) to reduce coupling and increase reusability. *SiloDriverIf* is the interface that the driver implements and provides to the controller. *SiloCtrl2DriverIf* is the required by the controller interface from the side of driver. This interface should be implemented by a controller that will interact with the specific Silo driver.

The properties of the real unit can be stored in the «cpController» or alternatively maybe encapsulated in another class of type «entity». The «driver», the «entity» and the «cpController» constitute the cyber part of the corresponding cyber-physical component.

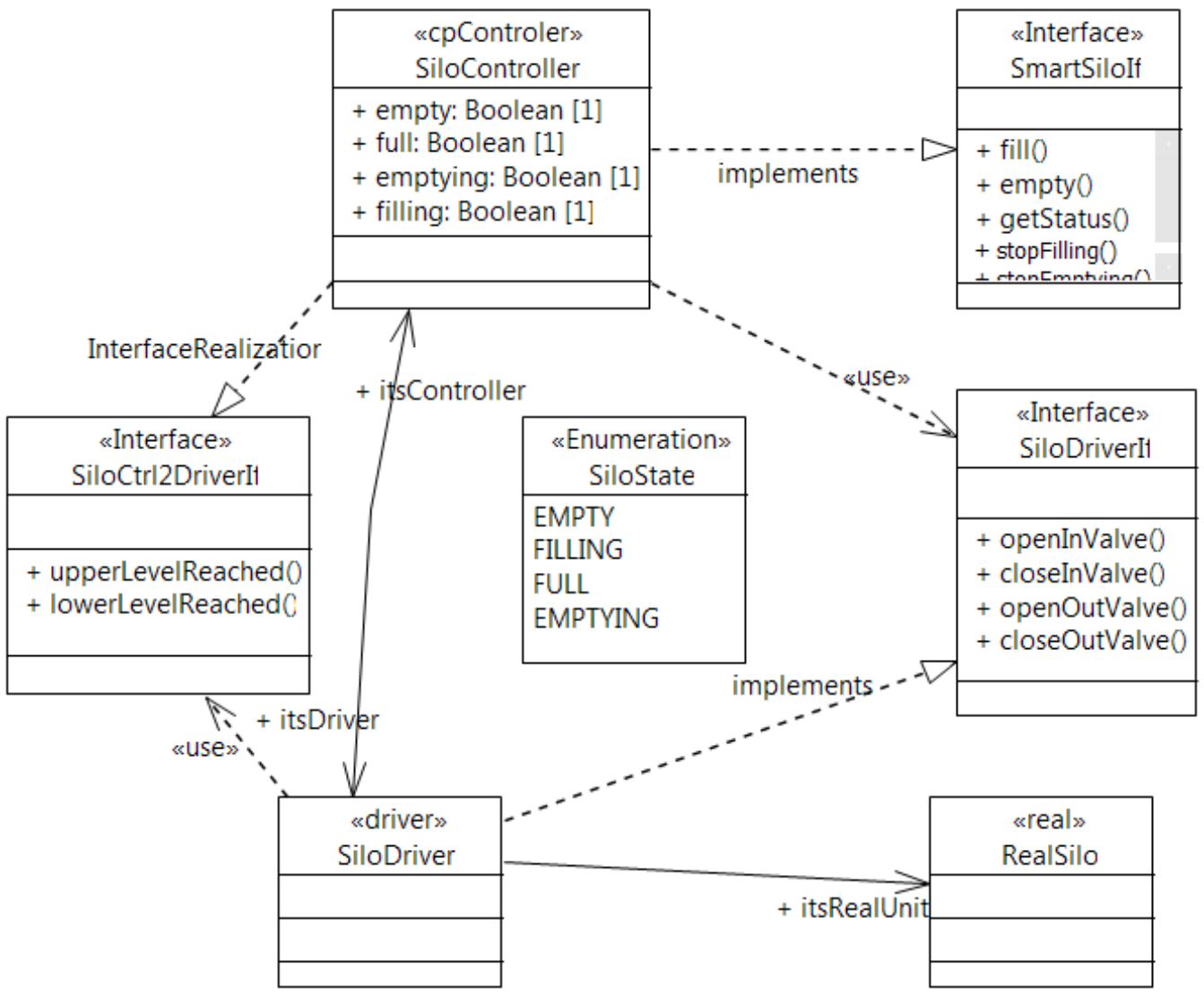

Figure 8. Capturing state, behavior and interfaces of the Silo «cyber» object.

Based on this modeling, fill() and empty() are operations of the Silo «cpController» class and FILLING and EMPTYING are states of the objects specified by this class. In this case, the state machine that is commonly used to represent the dynamic behavior of the real silo considers fill and empty as events from the environment. These events trigger a behavior that is depended on the object's state which maybe empty, full, filling or emptying. We do not use the inBetween state for simplicity reasons. We use the FB body to implement the event dispatching and the calling of the corresponding FB methods that have been defined to capture the behavior of the silo to the external events. Thus, the fill FB method is executed from the FB body in response to the fill event only when the object's state is empty.

Figure 9(a) presents the statechart specification of the dynamic behavior of Silo «cyber» that is realized in the case of composite object by the SiloController object. The initial state is a transitory state to activate the proper transition based on the initial status of the physical silo. In contrast to the approach presented in [10] an object is used to capture both the state and the dynamic behavior of the object. The OO concept is applied at the level of the physical object (silo) and effectively uses the benefits of this paradigm to increase reusability at this level. It should be mentioned that the Silo «cyber» may also be implemented as one object. In any case the implementation of the Silo «cyber» class is a decision of the developer.

FBs that capture the control logic of the processes are modeled using the «processController» stereotype and expose only cyber interfaces. Cyber components of type «processController» can be hosted in cyber-physical components that offer hosting services of the required QoS, if any, or they may be allocated to shared or exclusively used execution environments.

For the specification of «processController» classes we use UML 2.0 state machines or UML 2.0 activity diagrams. UML, which has accepted a broader use in the software industry compared to Petri nets, provides a more feasible and acceptable notation for modeling, compared to Petri Nets. Moreover, the adoption of UML 2.0 state machines, which are an object based variant of Harel statecharts and the UML 2.0 activity diagrams, which support modeling similar to traditional Petri nets, bring the benefits of formalism of these notations to the simplicity of use of UML. Figure 9(c) presents the state machine for the *GenLiqueurA* process controller. The exclusive use of pipe and power is obtained by modeling the common resource as shown in figure 9(b). Statecharts as well as activity diagrams of UML can be automatically translated to Petri nets for model checking [42][43]. However, the designer may select to model «ProcessController» classes directly in PNs as described in [10]. In this case an automatic transformation from PN to IEC 61131 is required since this transformation is not supported in [10]. This automatic transformation is a prerequisite to ensure that the 61131 implementation matches the PN specification.

For the mapping of the UML design specs to IEC 61131 specification we consider the following two approaches:

*1. Use FBD as target language for the transformation*

As argued in [19], the FBD provides limited support for OO and has introduced in the industrial automation domain a few basic concepts of the MDE. However, FBD provides only one kind of diagram that can be used to construct the model of the application software. This diagram is composed of FB instances and their interconnections.

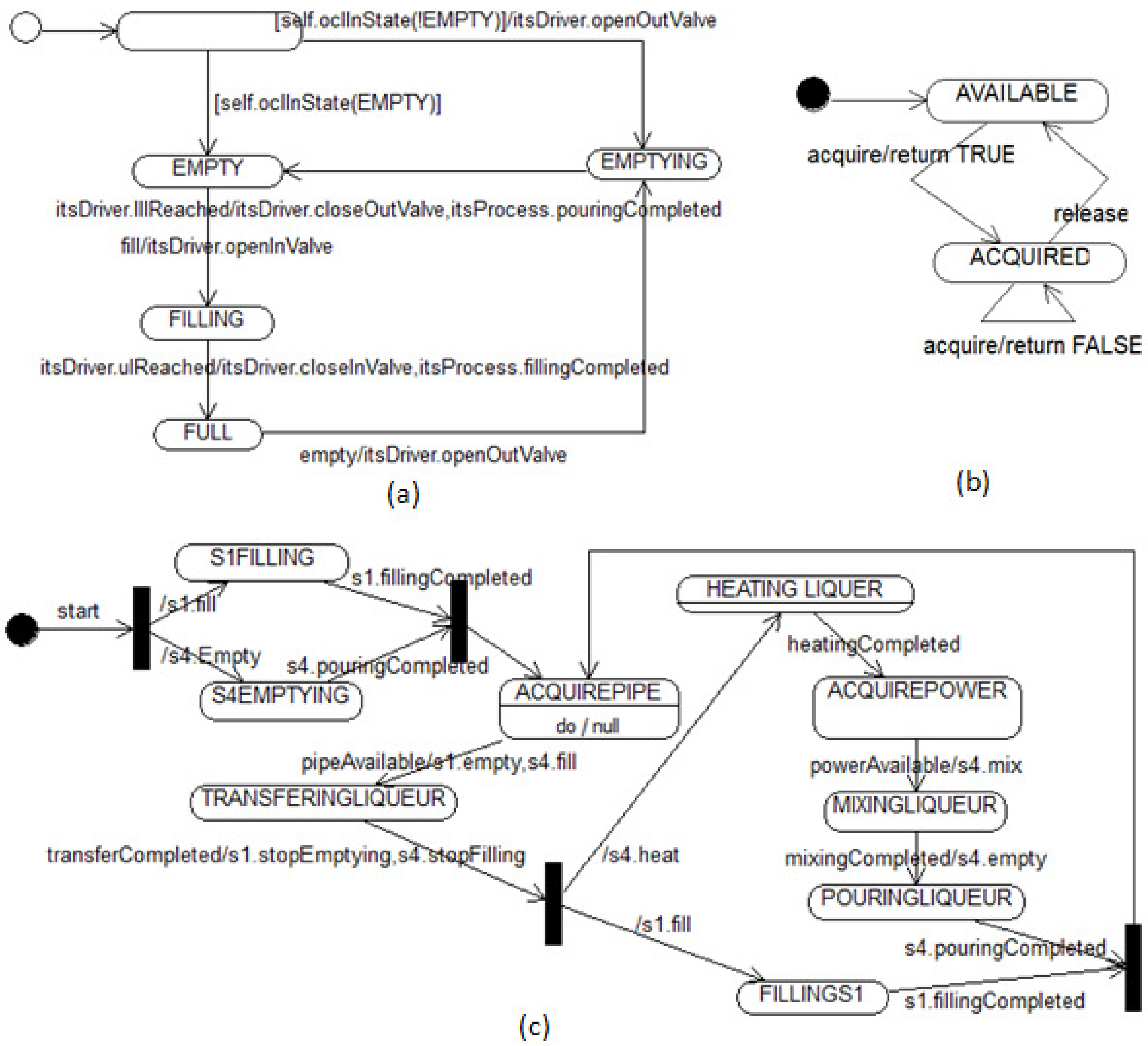

Fig. 9. State machines for (a) SiloController, (b) common resource and (c) *GenLiqueurA* cyber components.

*2. Use textual languages such as ST as target language of the transformation.*

The SysML/UML design specs are directly transformed to IEC 61131 textual languages. FBD is not used in this approach thus it does not impose its restriction in behavior but also in structure specification of the target model. A limited use of FBD just for interface specification may be adopted but in this case an enhanced OO FBD is required to support a coherent interface specification as is argued in [18].

For an effective use of FBD as target in the proposed MDE approach, additional diagrams should be defined to allow: a) more abstract models to be constructed, and b) more aspects of the system to be captured in order to have a complete and comprehensible model of the system [19]. But even in this case the benefits of using an extended FBD as target are not clear. This is why we have adopted the 2nd approach.

## V. Discussion

### *A. Modeling the Cyber part*

Following a system based approach, we define at the system layer, a) cyber components to capture the plant process logic, and b) cyber-physical objects to encapsulate physical objects and capture in their cyber parts the control logic that is required by the physical objects so as to transform these into smart components. We use the «processController» stereotype to model cyber components that capture the plant processes' logic and the «driver» and the «process-Controller» stereotypes to model the cyber part of cyber-physical objects.

In this section we discuss this approach in relation with the one presented in [10]. Based on this discussion we consider the integration of the two approaches.

Authors in [10] propose the use of DFBs and OFBs based on the following reasoning. They claim that the program of sequences, that is one of the fundamental things in industrial automation, “cannot be coded as a method since methods have local variables but they do not store data between two consecutive calls.” However, this is not true for OOP. Methods have local variables but they affect the state of the object on which they operate. Instance data members are used to represent the state of the object, while operations are used to capture the object’s behavior in response to external events. The execution of operations in response to external events is depended on its state and their execution usually affect its state. Authors claim that by using the concept of OFB an operation can be simply added by defining a new OFB class just for the new operation. However, they define the silo with mixer and heater (SILO_WITH_MIXER_AND_HEATER) to EXTEND the simple silo (SILO) so as to inherit its operations and add new ones.

The decision of authors in [10] to implement operations such as filling and emptying as classes (FILLING OFB and EMPTYING OFD) violates the basic principle of encapsulating state and behavior by using the construct of class. This approach is more close to the procedural programming paradigm where functions (OFBs) are called by passing the pointer of the data structure they operate on. In this way the state of silo is captured in the SILO DFB, while operations that affect the silo state and their execution is depended on the object’s state are captured in different classes. Thus, the dynamics of the silo object are distributed into two or more classes, as shown in figure 7. This design decision increases the coupling among structural components and reduces their cohesion. It makes the design more complicated and difficult to prove its correctness, but the most important is that it negatively affects reusability. Moreover, this approach is error prone since operations which are not related with a specific DFB may be applied on it. For example, the developer may connect a HEATING OFB with the SILO DFB. Moreover, authors claim that “the distinction between DFB and OFB helps to reuse DFBs which have direct interface with the field.” However, it is not clear why the filling of the silo behavior has no direct interface or relations with the field.

Instead of the DFB we define the driver of the silo, i.e., the proxy of the real world object in the cyber world shown in figure 8. This driver encapsulates the implementation details of the interaction of the SiloController (cyber world) with the real silo (physical world). It is common for the cyber world designers to extend the behavior of the physical world objects by constructing cyber world objects that provide higher level functionality, e.g., fill() and empty(), compared to the physical world objects functionality, e.g., openINValve() and closeINValve(). We capture this extra functionality along with the state of the physical object in the corresponding «cpController» object. Thus the state and behavior of smart silo is captured in the SiloController «cpController» class. This increases modularity, reusability and the consistency in the silo’s behavior. The «cpController» class is mapped to an OO FB. For example, for the silo we will have the SiloControllerFB. The body of this FB will implement the state machine shown in figure 9(a). It will call the corresponding method depending on its state. In this case methods are private and only the body of the FB is public. An alternative is to have the methods, which implement the operations on the silo, public and implement the state machine by a private method. Public methods will only perform a method call of the state machine private method with the proper argument. To implement this alternative the «cpController» should be mapped to an IEC 61131 class.

Authors in [10] use a PN controller to model the two processes that are performed by the plant. They model each process using a Petri Net. Thus the PN controller consists of two independent Petri Nets. To address the constraints imposed by the use of pipe and power they identify the need for coordination. They capture this coordination logic in what they call Supervisor and model it using Petri nets. The result of this approach is a Petri net that captures two service sequences, i.e., the two physical processes which are executed concurrently, and the coordination logic required for the exclusive use of common resources, i.e., the pipe and the power. Authors next implement this Petri Net in the main loop of the PLC program.

Our approach for the modeling of the two processes is different. We model each process separately using statechart or activity diagram. To address the requirement for exclusive use of the common resources pipe and power, we discriminate between two different implementations both in the context of the scan cycle model. If both cyber processes are executed by the same thread of control, as is the case with the implementation described in [10], then a method is used to check the availability of the common resource. The common recourse is not available when in state ACQUIRED as shown in figure 9(b). In this case, an action should be added to the statechart of the process *GenLiqueurA* to release the resource as soon as it is not required. In case of concurrent execution of «Cyber» processes, which is meaningless if both threads are executed in the same scan cycle, the common resource can be implemented as a monitor.

### *B. Support for distributed applications*

Authors in [10] claim that the languages of the IEC 61131 standard are not ready to meet the new challenges of widely distributed automation systems. They also claim that in the supervisory control theory, methodologies based on formal models have been developed to improve the coordination of concurrent and distributed systems. More specifically authors claim that they use supervisory control, in the context of industrial control, to solve the problem of programming concurrent FB behaviors, since as they claim a PLC program could be represented as a network of FBs that run concurrently to execute desired services’ sequences. However, their approach and mainly their example application through which they describe their approach is not distributed not even contain concurrently executed computational behaviors; the example application is developed based on the scan cycle model that is executed on one PLC as a sequence of FB calls inside a cycle. More specifically the example application, i.e., the program SILOS CORDINATION, which implements the PN controllers and the PN supervisor, is implemented as an execution of a sequence of FB calls inside a scan cycle. Thus, the benefits of the presented approach as far as it regards distributed applications and concurrently executed computational behaviors are not evident or at least not presented by the example application that is used to demonstrate the proposed approach.

Our approach that introduces the decomposition of the cyber physical system at the system layer in terms of cyber-physical and cyber components and adopts the IoT as the glue for the cyber interfaces provides inherent support for distribution.

### *C. The event driven execution and the exploitation of the service oriented paradigm*

POUs, in IEC 61131, may be executed either periodically (time triggered control) or upon the occurrence of a specific event (event-triggered control) even though this latter is not supported by all 61131 programming environments as also admitted in [10]. The absence of explicit control on the execution order of IEC 61131 FBs in an application is considered as a drawback of this standard [44]. However, as it is argued in [15] the 61131 languages allow for the explicit definition of the execution order of FB instances, with the only exception being the graphical representation of the FBD. But even in this case, all commercially available tools address this problem by allowing the developer to explicitly specify the execution order.

Moreover, there is a big misperception regarding the event driven model introduced by IEC 61499. This misperception is evident, as claimed in [15], by examining the IEC 61499 Compliance Profile for Execution models. This profile, which defines according to [10, Ref. 13] an execution model, identifies the need to "have a predefined order before execution, such that during the course of a scan, all FB's within an FBN will always be invoked in the same order" [45]. This is obtained by having the designer to assign a unique priority to each FB of the FBN. This priority is actually defining the execution order of the FB instances during every scan cycle. It should be noted that all this discussion refers only to the FBD since in ST, the one used for implementing the PN controller and the supervisor in [10], the execution sequence is clearly defined.

Furthermore, as argued in [15], it is not clear if the term *event driven* in IEC 61499 refers to the handling of external events or to the handling of the internal ones. Most of the IEC 61499 journal papers consider it as referring to the handling of the internal events. Others relate it to the handling of external ones. Both categories refer the relation or mapping of the event driven to the scan based [15]. It is considered that the objective of the event interface is the explicit definition of the execution sequence of FB instances in a Function Block Network (FBN).

Authors in [10] claim that they present an event-driven approach to improve the design of industrial control systems using commercial PLCs. However, the design they propose and the subsequent implementation they present for the example application are based: a) on the scan cycle model, i.e., time triggered control, that is executed on one PLC, and b) on the execution of sequences of computational processes based on the method call paradigm (FB calling). Moreover, authors claim that FBs are seen as service providers and event inputs are seen as service requests. Based on this view it is claimed that the presented approach exploits the service-oriented paradigm. Authors also claim that one of the benefits of the presented FB model is that a service (i.e., the associated algorithm) is executed only when it is explicitly invoked by means of the associated event, but they do not explain what is the benefit of this when the control application is based on the scan cycle model as is the case for their proposal. Authors consider each event input as a service request and they propose to implement the expected by the FB behavior as an algorithm. However, if we consider the event INIT of the Silo DFB shown in [10, Fig. 5] and the corresponding INITOK, it is assumed, by applying the event-based execution order, that INITOK of one silo DFB will be connected to the INIT input of another to specify the execution order of the initialize algorithms captured in the silo DFBs. In this case, it appears that the initialize service of the second and subsequent silo DFBs is requested by the previous silo DFB, which is not in harmony with the semantics of the application and the service oriented computing. Moreover, there are no methods or inputs (event or data) in the silo DFB [10, Fig. 5] to trigger the behaviors related to open and close valves, which should be captured by the silo DFB, even though this behavior is captured by the silo FBD as claimed in [10, p.995]. This is also not shown in the textual representation of the silo DFB.

Our approach, which is based on clearly defined cyber physical components that offer their services through IoT at the system layer, provides inherent support for the exploitation of the service oriented paradigm.

From the above it is evident that an integration of the two approaches, the one presented in [10] and the one presented in this paper, can be considered as far as it regards the modeling of the cyber part and more specifically the use of Petri Nets and Supervisor theory. This may help in avoiding the transformation of UML design specs to Petri nets but on the other side an automatic transformation of Petri nets to IEC 61131 code is required. Petri nets may also be used for service composition [46]. In addition, some implementations details on IEC 61131 presented in [10] can also be utilized for the transformation of UML/SysML designs to IEC61131 code.

## VI. CONCLUSION

We have presented a system engineering approach for the modeling and development of industrial automation systems based on the new version of the IEC 61131 standard for systems running on PLCs. This approach adopts the 3+1 SysML-view model and extends it to match the requirements of the industrial automation systems domain. The decomposition of the system at the system layer in terms of cyber-physical and cyber components and the adoption of IoT as a glue regarding the cyber interfaces provides an inherent support for distribution and exploitation of the service oriented and component based paradigms. Our approach also exploit the MDE paradigm through the definition of a domain modeling language by use of UML/SysML profiles. However, an automatic transformation of the UML/SysML design specs to IEC 61131 is required. We are working on such a transformation so as to fully exploit the benefits of MDE. We have discussed a similar approach presented in [10] and considered the integration of both approaches to form a more rigorous development process for IASs. We have used the example system to demonstrate the applicability of the presented approach.